\documentclass[12pt,a4paper]{article}
\usepackage{graphicx,amssymb,amsmath}
\usepackage{times,mathptm}
\usepackage{epsfig}
\usepackage{color}
\usepackage{float}
\restylefloat{figure}

\begin{document}

\title{Effects of Sudden Changes in Inflow Conditions on the Angle of Attack on HAWT Blades}

\author{B. Stoevesandt\thanks{Correspondence to: B. Stoevesandt or J. Peinke, ForWind - Center for Wind Energy Research, Institute of Physics, Carl von Ossietzky University of Oldenburg, D-26111 Oldenburg, Germany. E-Mail: bernhard.stoevesandt@uni-oldenburg.de} and J. Peinke\\
 \small{ForWind - Center for Wind Energy Research}\\
 \small{Institute of Physics, Carl-von-Ossietzky University of Oldenburg}\\
 \small{D-26111 Oldenburg, Germany}}
 
\date{}
\maketitle

\begin{abstract}
In this paper changes in wind speed and wind direction from a measured wind field are being analyzed  at high frequencies. This is used to estimate changes in the angle of attack (AOA) on a blade segment over short time periods for different estimated turbine concepts.  Here a statistical approach is chosen to grasp the characteristics of the probability distributions to give an over all view of the magnitude and rate of the changes.
The main interest is the generation of basic distributions for the calculation of dynamic stall effects and stall flutter due to wind fluctuations.
\end{abstract}

%
%
\section{Introduction}
Dynamic stall is a concept to calculate strong, yet hard to predict, load fluctuations on wind turbines. One of the main causes for dynamic stall and stall flutter on wind turbines are sudden changes in wind speed and wind direction leading to changes in the angle of attack (AOA) on wind turbine blades. Both angle of attack and dynamic stall are part of the 2D calculation concept used by most blade element moment (BEM) or lifting line model theory programs to calculate the main aerodynamic properties for wind turbines \cite{breton1}\cite{snel1}. Dynamic stall is mostly taken into account by empirical models based on the knowledge of the amplitude and fequency of the changes in the angle of attack \cite{mccroskey1}.\\
Most of the research on the dynamic stall models evolved from helicopter research like the ONERA model \cite{tran1} or the Beddoes-Leishman model \cite{leishman1} and were later transfered to wind turbine applications. At helicopters dynamic stall mainly appears due to a rapid change in AOA induced by a - from the rotor perspective - yawed inflow \cite{reber1} caused by the flight velocity. The changes in AOA and its rates of change are periodic in time and thus can easily be calculated. Such change rates are also often used for wind turbine dynamic stall modeling \cite{gupta1}\cite{gonz1}.\\
In case of a natural wind field the situation is different, as changes in the AOA are induced by the incident wind field which appears in most cases to be highly turbulent. To calculate the amplitudes and frequencies of changes in AOA the distributions of changes in wind direction and wind speed have to be investigated.\\
The fluctuations of the wind has attracted attention in research in recent years (e.g. \cite{sura03}\cite{monahan06}\cite{boettcher07}). The investigation on the influence of such fluctuations on the AOA is a relevant consequence. Bierbooms and Veldkamp did some research on wind speed fluctuations and the resulting loads on wind turbines \cite{bierbooms04}\cite{bierbooms07}.\\
For the load estimation on wind turbines, wind field generators using a wind field based on a Gaussian distributed wind speed fluctuations are applied quite often (e.g. \cite{fogle1}\cite{jonkman1}\cite{natarajan1}\cite{moriarty1}). This Gaussian approach leads to an underestimation of extreme values of the fluctuations in wind speeds (see e.g. \cite{boettcherphd}).\\
Changes in wind direction lead to a yawed inflow condition, which is one reason for wind turbines to face dynamic stall effects \cite{schreck1}\cite{schreck2}.The analysis of wind directional changes has been described by van Doorn et al. \cite{doorn1}. The methods are picked up in this paper and applied to a measured wind field for further statistical analysis.\\
If the fluctuations in wind speed and changes in wind direction would be strongly correlated, it would be straight forward to develop a comprehensive analytical model for the fluctuations in the AOA. However this does not often seem to be the case \cite{hansen1}. Therefore we propose a direct calculation of the fluctuations in the AOA from changes in both, the wind speed and the wind direction.\\
Due to the unsteadiness of the wind field and the response of the wind turbine, the actual AOA is hard to determine. Different approaches to resolve this problem have been proposed \cite{shen1}\cite{shen2}\cite{sant1}\cite{sant2}.
In this paper we assume that the actual AOA is unknown. Yet the changes in the AOA due to changes in wind speed and wind direction are evaluated. We do so by estimating the changes in the AOA over short period of time. Therefore will assume certain simplified models for such changes on an arbitrary chosen position on a rotating blade.\\
Here we will explain the methodology to calculate the local changes in the AOA, using a measured wind field as input. At first a short analysis of the wind field data will be done. Next, for two different wind turbine modeling principles, the changes in AOA are calculated and analyzed. Included is an analysis of the changes in AOA only due to changes in wind direction. In the end an analysis of the time scales on which the changes appear and the characteristics of their distribution are presented with the aim to improve the investigation of dynamic stall effects and their modeling for wind turbines.\\

%
\section{Characteristics of the measured wind}
\label{Measurement}
Before we present a method to calculate the changes in AOA, the measured wind field we used for the demonstration shall be discussed here on the bases of its statistical aspects.\\
The data from a measurement campaign at a wind farm at Meerhof near Paderborn in Germany is being used throughout this paper \cite{gottschallphd}. The data was measured using a Gill R3-50 ultra sonic anemometer in 98m height with a sampling rate of 50 Hz. If not otherwise stated, it has been averaged to a 10 Hz sampling rate. Using the Taylors frozen turbulence hypothesis this sampling rate corresponds to a length scale of 1 m at a wind speed of $10$ m/s. Wind coming from the sector from $40.5^\circ$ to $133.5^\circ$ has been excluded from the analysis, since a wind farm was near in these wind directions. This way an influence by the wake of the wind farm on the wind measurement could be avoided. For the investigation a dataset of seven consecutive days from December 1st to 7th, 2005 has been used, leading to a dataset of $6\cdot10^6$ samples at 10 Hz.\\
 
\begin{figure}[htbp]
\begin{center}
\epsfxsize=3.0in
\epsffile{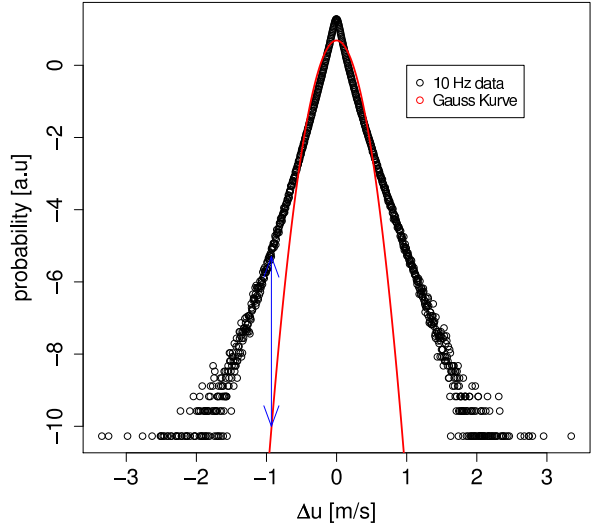}
\end{center}
\caption{Probability density function (pdf) of all wind speed increments for $\frac{1}{\tau}=10$ Hz. The typical intermittent behavior is seen by the strong deviations of the tails from a fitted Gaussian distribution. Here like for all other plots we used only wind speeds at $|u| \geq 2$ m/s.}
\label{windspeed}
\end{figure}

%
%
%
%
To give an impression of the wind field used in this study we describe different relevant characteristics of the wind field and the methods for the characterization of turbulent fluctuations.\\
The mean measured wind speed over all data was $\bar u=6.96$ m/s with a turbulence intensity of $I=13.3$\% according to IEC-61400-1 Standard \cite{iec61400}, calculated from the average standard deviation of 10 minutes periods.\\
For the further calculations some approximations have been made. We exclude wind speeds of $|u| < 2$ m/s in the analysis. Changes in wind direction at low wind speeds are very sensitive to measurement errors. Very low wind speeds are also irrelevant for wind turbines, due to the lack of induced relevant loads.\\
Since the dataset consisted only of one point measurements, we will also make the assumption for the calculations that the wind field and its turbulence is homogeneous. Even if this is in reality not the case, implementing a different wind field on the method will be straight forward.\\
For the research of dynamic stall we are mainly interested in the extend of the changes at certain periods. Therefore the wind field and its turbulent structure was further analyzed using incremental statistics. They are a good measure for fluctuations at steps $\tau$.
Increments of a property $\xi(x,t)$ are defined as
 \begin{eqnarray}
 \xi_{\tau}(x)= \xi(x,t+\tau)-\xi(x,t)\;\;. \label{Increment}
 \end{eqnarray}
where $\tau$ is the time scale to be considered. The statistics of the increments allow a time scale resolved analysis of the fluctuations of a magnitude $\xi$.\\
For the measured data the probability density function (pdf) of the wind speed increments for a time scale of $\tau$=10 Hz is given in fig. \ref{windspeed} in semi-logarithmic presentation. We observe a typical intermittent behavior of the pdfs, which has been described e.g. by B\"ottcher \cite{boettcherphd}. The distribution deviates strongly from a fitted Gaussian curve. In our case the empirical probability of wind speed changes at 10 Hz and of magnitude 0.85 m/s is already by a factor $\ge 10^4$ higher than a corresponding Gaussian distribution would predict. For larger increments this deviance increases by further orders of magnitude.\\

%
%
\begin{figure}[htbp]
\begin{center}
$\begin{array}{c@{\hspace{0.3in}}c}  
 \multicolumn{1}{l}{\mbox{\bf }}  &
 \multicolumn{1}{l}{\mbox{\bf }} \\ [-0.5cm]
\epsfxsize=2.8in
\epsffile{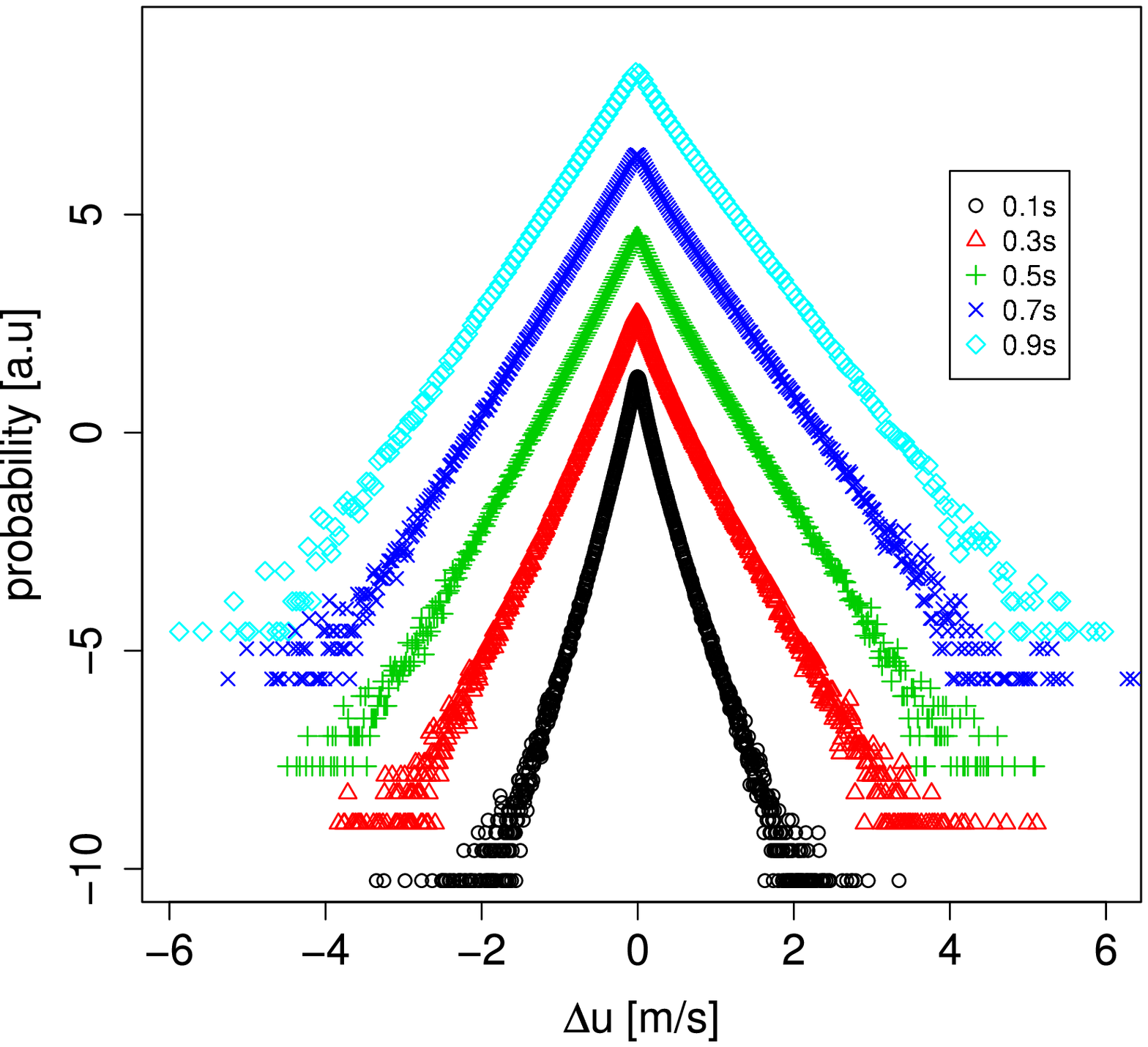}
 \put(-200,150){a)}  & 
\epsfxsize=2.8in
\epsffile{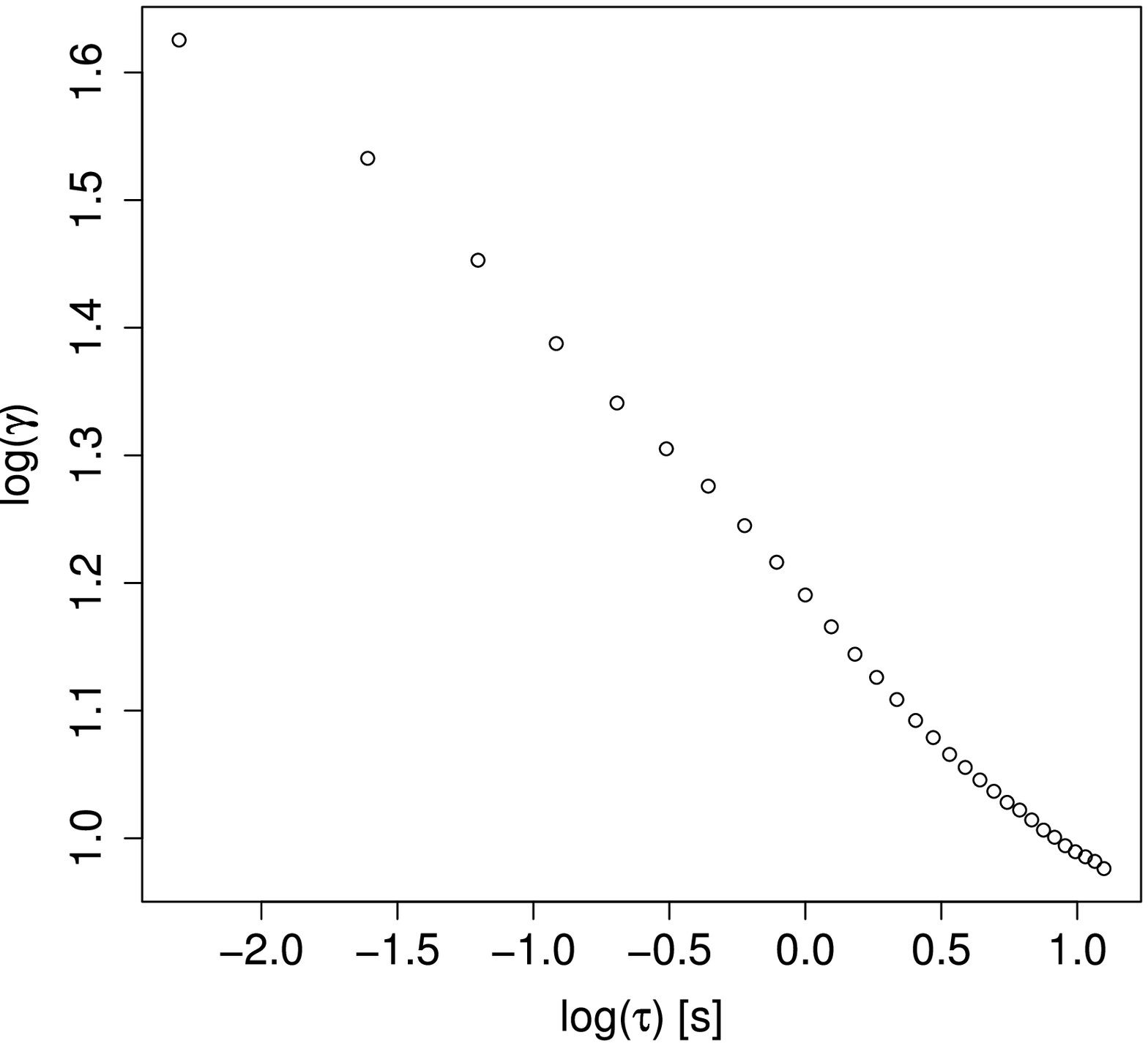}
\put(-200,150){b)}\\ [1.0cm]
\multicolumn{1}{l}{\mbox{\bf }}  &
\multicolumn{1}{l}{\mbox{\bf }}   \\ [-1cm]
\end{array}$
\caption{pdfs wind speed increments for different time scales $\tau$ on a semi-logarithmic scale a). The values of the excess kurtosis of the pdfs give an idea of the magnitude of the deviation from Gaussian distributed pdfs with a $\gamma=0$.}
\label{kur-wind}
\end{center}
\end{figure}

%
%
The deviance from a Gaussian curve can be quantified by the kurtosis and the skewness, where the kurtosis can be used to describe a symmetric deviation from a Gaussian curve and the skewness to describe the asymmetry of the curve. In this paper the excess kurtosis is taken into account given by\\
\begin{eqnarray}
\gamma=\frac{\frac{1}{n}\sum_{i=1}^{n}(x_i - \overline{x})^4}{(\frac{1}{n}\sum_{i=1}^{n}(x_i-\overline{x})^2)^2}-3\;\;. \label{kurt} 
\end{eqnarray}
The excess kurtosis is normed so that $\gamma=0$ for a Gaussian distribution and $\gamma>0$ for intermittent pdfs.\\
The skewness is calculated by the standardized third moment:\\
\begin{eqnarray}
\nu=\frac{\frac{1}{n}\sum_{i=1}^{n}(x_i - \overline{x})^3}{(\frac{1}{n}\sum_{i=1}^{n}(x_i-\overline{x})^2)^{\frac{3}{2}}}\;\;. \label{skew} 
\end{eqnarray}
It follows that $\nu=0$ for symmetrical pdfs.\\
For a characterization of the wind field, we can determine the pdfs of the wind speed increments at different time scales $\tau$ (see fig. \ref{kur-wind}).  In fig. \ref{kur-wind} b) the corresponding values of the excess kurtoses are shown as a function of time scales between 0.1s and 3s is shown. With an increase in time period of wind speed increments $\gamma$ becomes smaller, which means that the shape of the pdfs approach slowly the shape of a Gaussian curve. A linear regression indicates a characteristic $\gamma \approx \tau^{-0.21}$. An extrapolation would lead to a kurtosis of a Gaussian curve at a time scale of 150 s. This is however a time scale far above the scales at which most control systems adapt wind turbines to the surrounding conditions. Thus it is of no further interest for our considerations here.\\

\begin{figure}[htbp]
\begin{center}
$\begin{array}{c@{\hspace{0.3in}}c}  
 \multicolumn{1}{l}{\mbox{\bf }}  &
 \multicolumn{1}{l}{\mbox{\bf }} \\ [-0.5cm]
\epsfxsize=2.7in
\epsffile{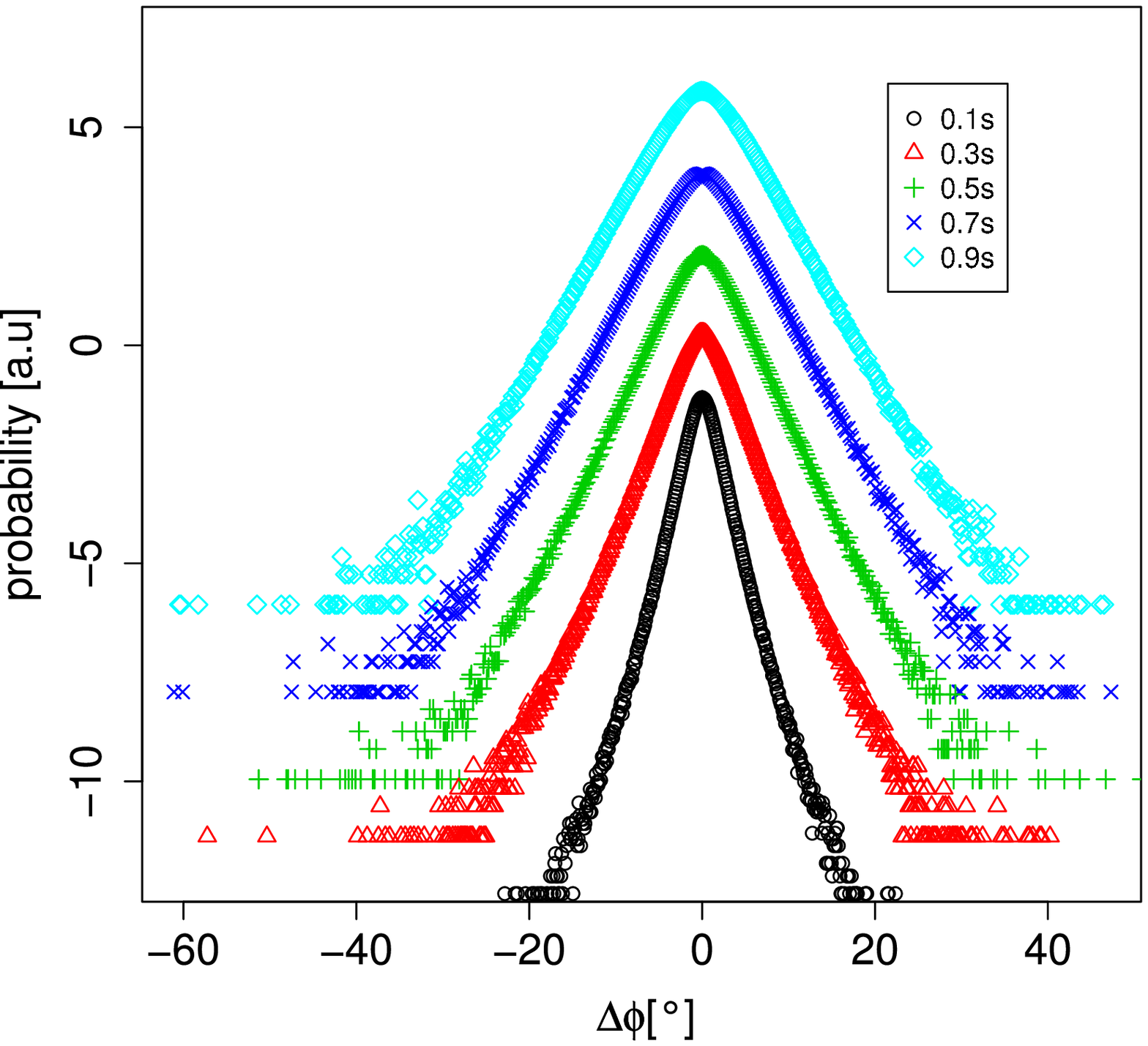}
 \put(-200,150){a)}  & 
\epsfxsize=2.7in
\epsffile{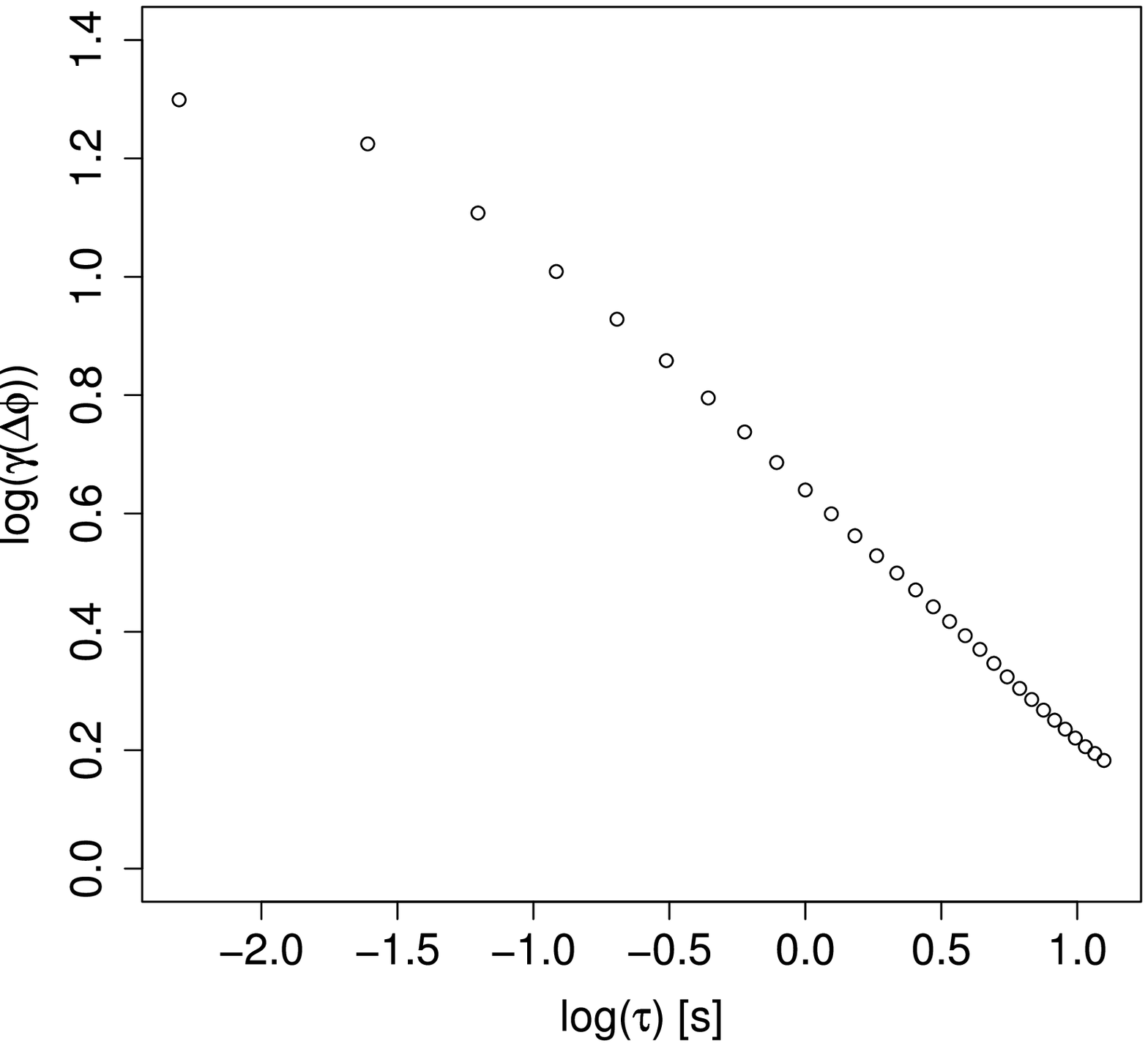}
\put(-200,150){b)}\\ [1.0cm]
\multicolumn{1}{l}{\mbox{\bf }}  &
\multicolumn{1}{l}{\mbox{\bf }}   \\ [-1cm]
\end{array}$
\end{center}
\caption{a) probability density functions of wind direction increments at different time scales on a semi-logarithmic scale and b) the excess kurtoses.}
\label{winddir}
\end{figure}

So far we have look just at the wind speed $|u|$ and its increments. For our purpose it is also important to take the wind direction $\phi$ and its increments into account.\\
The pdfs of the increments of the wind direction show again an intermittent form (see fig. \ref{winddir}). The values of the excess kurtoses are lower than for the wind speeds.
%
%
Fig. \ref{winddir} makes clear, that wind direction changes are large and appear rapidly. Changes of $10^\circ$ and more within $0.1$ seconds do occur. For larger time scales, still less 1 second, changes of up to $20^\circ$ occur. The double logarithmic presentation of the excess kurtoses in fig. \ref{winddir} b) leads to a similar power law with $\gamma\approx \tau^{-0.42}$ for scales $0.5 \leq \tau \leq 3.0$. An interpolation leads to a kurtosis of a Gaussian curve at $\tau \approx 4.6$ s.\\

As both changes in wind speed and in wind direction can cause changes of the angle of attack, the correlation of the appearances of both phenomena is of particular interest. A possible measure to quantify this is the so called gust directional index (GDI) 
\begin{eqnarray}
 GDI(\tau)=\frac{abs(u(t+\tau)-u(t))}{max(abs(u(t+\tau)-u(t)))}+\frac{abs(\phi(t+\tau)-\phi(t))}{max(\phi(Dir(t+\tau)-\phi(t)))},
\label{GDI}
\end{eqnarray}
 as described by Hansen et al. \cite{hansen1}.
Here $u(t)$ is the wind speed at time $t$, $\tau$ is the time scale and $\phi(t)$ the wind directione. The $GDI \in [0,2]$ takes into account two simultaneous increments $u_{\tau}(t)$ and $\phi_{\tau}(t)$, with $GDI(\tau)=2$ if both magnitudes show extreme events occurring in a correlated way and $GDI(\tau)=0$ for complete non-correlation. For quantification the probability of strongly correlated events is regarded \cite{hansen1}.\\
For an 1 Hz sampling rate the probability of the $GDI \geq 1.98$ for $\tau=10$ s was at $1.3 \times 10{-5}$.  For a sampling rate of 10 Hz this value for $\tau=10$ s decreases to $1.5 \times10^{-7}$. It seems a large simultaneous change in wind speed and direction is not likely to appear, the numbers of large GDI events are comparatively low.\\
The low values indicate for the question of the changes in angle of attack, that changes in wind speed and direction could mostly be considered separately for this wind field. Nevertheless this does not seem to be the case for all wind situations (compare \cite{hansen1}).\\

\section{Analysis of fluctuations in the AOA}
\subsection{The calculation concept}
Instead of focussing on aspects of the wind field components separately we proceed to calculate the changes in AOA by a comprehensive approach. To do so a model of an estimated wind turbine is need. The real AOA on wind turbines is strongly dependent on its control system. Here we do not intend to go into details of a specific control system. Instead we consider two different simplified cases:
\begin{enumerate}
\item[a)] A classical fixed tip speed ratio turbine that adapts tip speed and pitch angle to the changes in the averaged wind of a certain period $\tau_a$  by a fast controlling system. We will take here $\tau_a = 2$ s.
\item[b)] A turbine with a fixed rotational speed, that adapts the pitch angle to the averaged wind of the last 2 seconds.
\end{enumerate}
The models will further be labeled as case a) or b) respectively.\\
For all cases of the models we find a finite time $\tau_a$ for the control system to pose a wind turbine in its optimal working condition. In case of wind changes on time scales $\tau \le \tau_a$ the control system is not able to react quickly enough. For these situations the fluctuations of the wind conditions e.g. grasped by increments are of importance. This aspect will be studied in the following.\\
 \begin{figure}[htbp]
\centering
\includegraphics[width=4.2in]{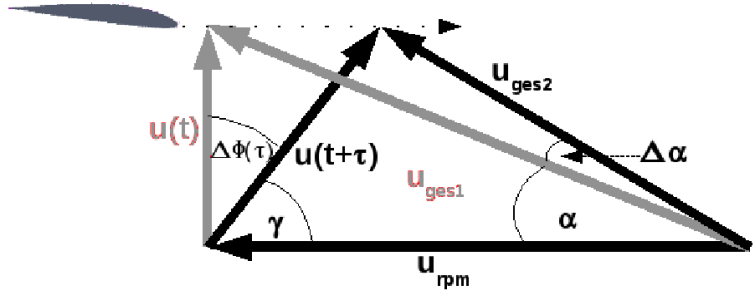}
\caption{The changing AOA can be calculated by the law of cosines. In grey is the original wind velocity and direction - black is the new one after a time $\tau$.}
\label{cosines}
\end{figure}
The used estimations of turbines enable us to determine a change in AOA by a change in wind situation. Such changes can then be calculated by using the geometry shown in fig. \ref{cosines}, where
\begin{eqnarray}
\Delta \phi=\phi(t+\tau)-\overline\phi(t). \label{sindt}
\end{eqnarray}
Here $\overline \phi(t)$ is the averaged angle between the incoming wind and the rotor plane determined by the ``slow'' yaw system. \\
The general equations for the changing AOA $\Delta \alpha$ in fluctuating winds are, in case a) of the fixed tip speed ratio turbine \\
\begin{eqnarray}
\Delta\alpha = \arccos \left( \frac { \overline u(t) \frac{\lambda_T r}{R} - \frac{2}{3}u(t+\tau)\sin(\Delta \phi)}{\sqrt{(\overline u(t) \frac{\lambda_T r}{R})^2+(\frac{2}{3}u(t+\tau))^2-\frac{4}{3}u(t+\tau) \overline u(t) \frac{\lambda_T r}{R}\sin(\Delta \phi)}}\right)-\arctan\left(\frac{2}{3}\frac{1}{\frac{\lambda_T r}{R}}\right). 
\label{tipspeedall}
\end{eqnarray}
For the fixed rotational speed turbine case b) we use\\
\begin{eqnarray}
u_{rpm}=\frac{2\pi r n_r}{60} \label{rpmeq}
\end{eqnarray}
where $n_r$ is the rotational speed in number of rounds per minute. For turbines with a rotor radius of $R=40$ m, typical values of $n_r$ lay between 10 and 20. Thus we obtain:
\begin{eqnarray}
\Delta\alpha = \arccos \left( \frac { u_{rpm}- \frac{2}{3}u(t+\tau)\sin(\Delta \phi)}{\sqrt{u_{rpm}^2+(\frac{2}{3}u(t+\tau))^2-\frac{4}{3}u(t+\tau) u_{rpm}\sin(\Delta \phi)}}\right)-\arccos\left(\frac{u_{rpm}}{\sqrt{u_{rpm}^2+(\frac{2}{3}\overline u(t))^2}}\right). \label{constspeedall}
\end{eqnarray}
Here in both cases $u(t)$ is the absolute value of the wind velocity at the time $t$, $\lambda_T$ is the tip speed ratio, which was taken to be 7 for case a). $R$ is the total rotor radius, which is considered to be $R=40$ m in this paper. $r$ is the position of the blade, which is to be regarded. The factor of $\frac {2}{3}$ in both equations reflects the estimated deceleration of the wind due to the blockage effect.\\
All calculations were done at a position of half of the blade length $r=20$ m. Although the absolute changes in AOA depend on $r$, the form of the distributions of the increments do not.\\

\subsection{Overall estimation of changes in AOA}
\begin{figure}[htbp]
\begin{center}
$\begin{array}{c@{\hspace{0.3in}}c}  
 \multicolumn{1}{l}{\mbox{\bf }}  &
 \multicolumn{1}{l}{\mbox{\bf }} \\ [-0.5cm]
\epsfxsize=2.7in
\epsffile{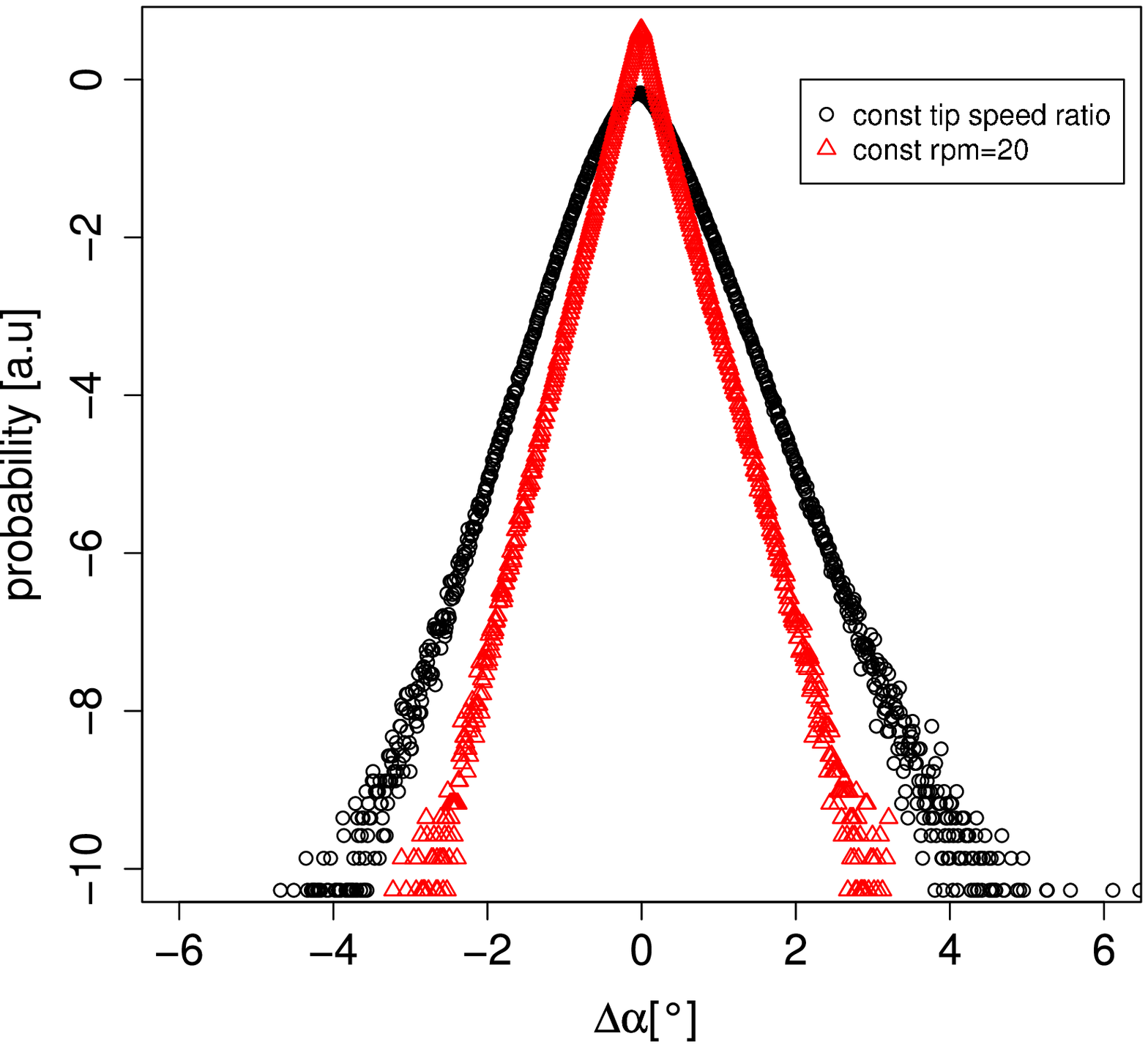}
 \put(-200,150){a)}  & 
\epsfxsize=2.75in
\epsffile{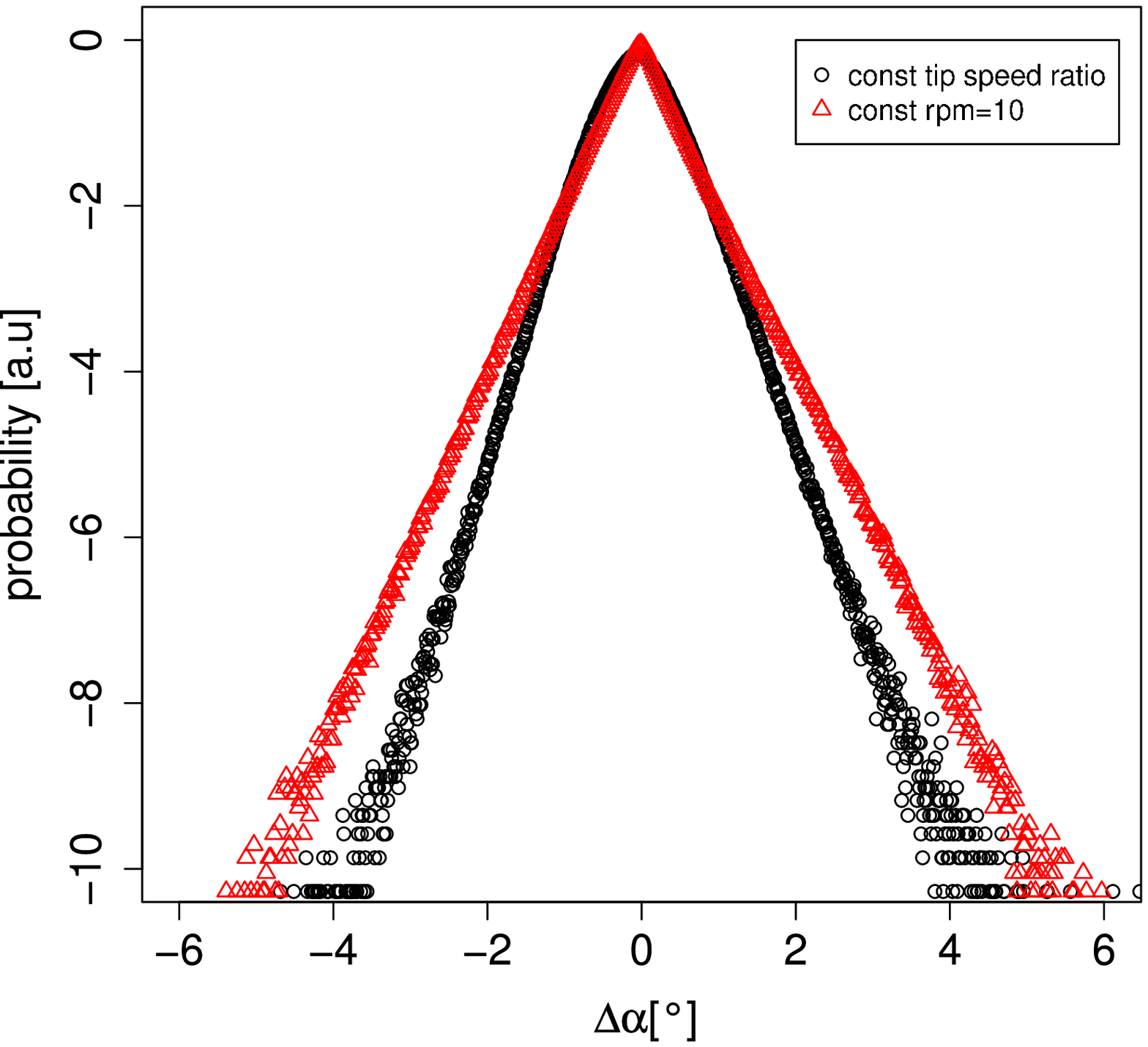}
\put(-200,150){b)}\\ [1.0cm]
\multicolumn{1}{l}{\mbox{\bf }}  &
\multicolumn{1}{l}{\mbox{\bf }}   \\ [-1cm]
\end{array}$
\end{center}
\caption{Pdfs of the changes in AOA due to change in wind over the period of $\tau_a=2 s$. In both figures a) and b) the changes of the calculations for constant rotational speed against (case b)) constant tip speed ratio (case a)) are given. In figure a) the const. rotational speed is given by a rotation of $n_{rpm}=20$, whereas in fig. b) $n_{rpm}=10$. Note that the pdf of case a) with depicted by the round symbols is the same in both plots.}
\label{aoages}
\end{figure}

In fig. \ref{windspeed} and fig. \ref{winddir} we can conclude that the timescales of rapid fluctuations do not differ very much between wind speed change and wind direction change. Further we can conclude from the low value of the GDI in section \ref{Measurement} that we cannot expect a strong correlation of the fluctuations in wind speed and wind direction. Therefore it makes sense to investigate the changes in the AOA by regarding the over all change in wind speed and direction at the same time, calculated by equations (\ref{tipspeedall}) and (\ref{constspeedall}).\\
Here the models for case a) (const. tip speed) and case b) (constant $u_{rpm}$) are taken into account. We use $\tau_a=2$ s. Fig. \ref{aoages} presents the pdfs of the increments $\Delta \alpha$ for both cases, in a) with $u_{rpm}=20$ in b) $u_{rpm}=10$ representing the limits of the range of rotational speed for a typical R=40 m turbine. To emphasize the difference in shape of the pdfs between the two cases a) and b) the results have been plotted together for each $u_{rpm}$.\\
The pdfs in fig. \ref{aoages} for both assumed models of the turbines show a strong intermittent structure. In case of the case a) turbine with constant tip speed ratio, the intermittency is less distinct. This is also quantified by the kurtoses of the pdfs in table \ref{tabaoages}. The kurtosis of the case a) pdf is  much lower yet the standard deviation $\sigma$ is relatively high compared to the pdfs of case b). The reason lies in the adaption of the case a) turbine rotation to the wind speed filtering some of the extreme fluctuations.\\
Further a slight asymmetry in the pdfs could be observed which are quantified by the skewness in  table \ref{tabaoages}. Since the fluctuations in wind speed should not cause such an asymmetry, this effect should be a result from the changes in wind direction. To quantify the influence of the changes in wind direction on the pdfs of the changes in angle of attack, we will proceed to analyze this effect separately.\\
\begin{table}
\centering
\begin{tabular}{|c|c|c|c|c|c|}
\hline
{\bf Type} & {\bf Skewness} & {\bf Kurtosis} & {\bf $\sigma$}\\
\hline
$\lambda_T=7$ & 0.20 & 1.8 & 0.59$^\circ$\\
$n_{rpm}=10$ & 0.16 & 3.1 &  0.71$^\circ$\\
$n_{rpm}=20$ & 0.17 & 3.2 &  0.36$^\circ$\\
\hline
\end{tabular}
\caption{Characterization of the pdfs for the change in AOA due to wind fluctuations - including wind speed and direction.}
\label{tabaoages}
\end{table}
 
\subsection{Maximum changes of attack due to sudden yaw angle changes}
\begin{figure}[htbp]
\begin{center}
$\begin{array}{c@{\hspace{0.3in}}c}  
 \multicolumn{1}{l}{\mbox{\bf }}  &
 \multicolumn{1}{l}{\mbox{\bf }} \\ [-0.5cm]
\epsfxsize=2.78in
\epsffile{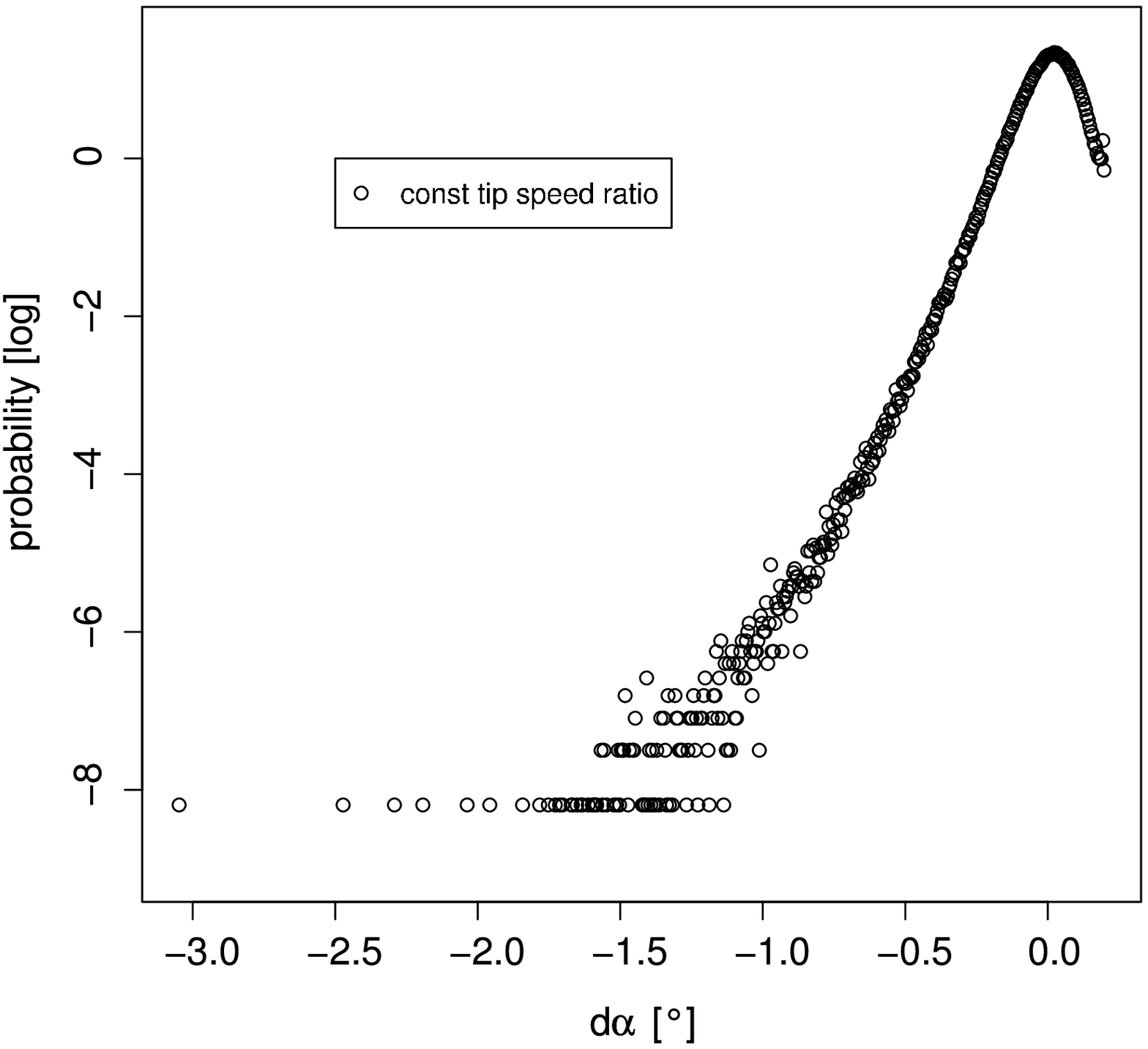}
 \put(-210,150){a)}  & 
\epsfxsize=2.78in
\epsffile{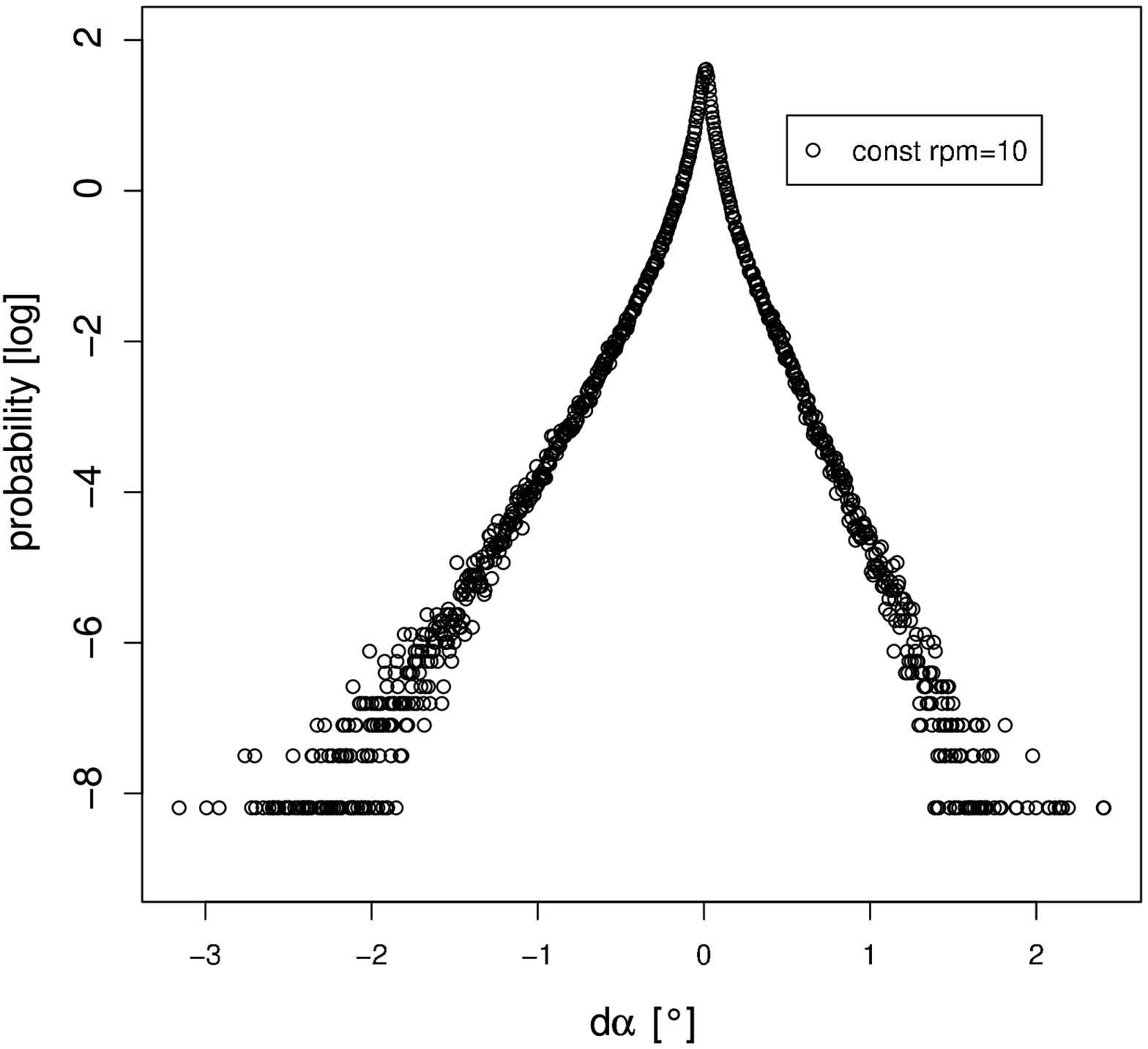}
\put(-200,150){b)}\\ [1.0cm]
\multicolumn{1}{l}{\mbox{\bf }}  &
\multicolumn{1}{l}{\mbox{\bf }}   \\ [-1cm]
\epsfxsize=2.78in
\epsffile{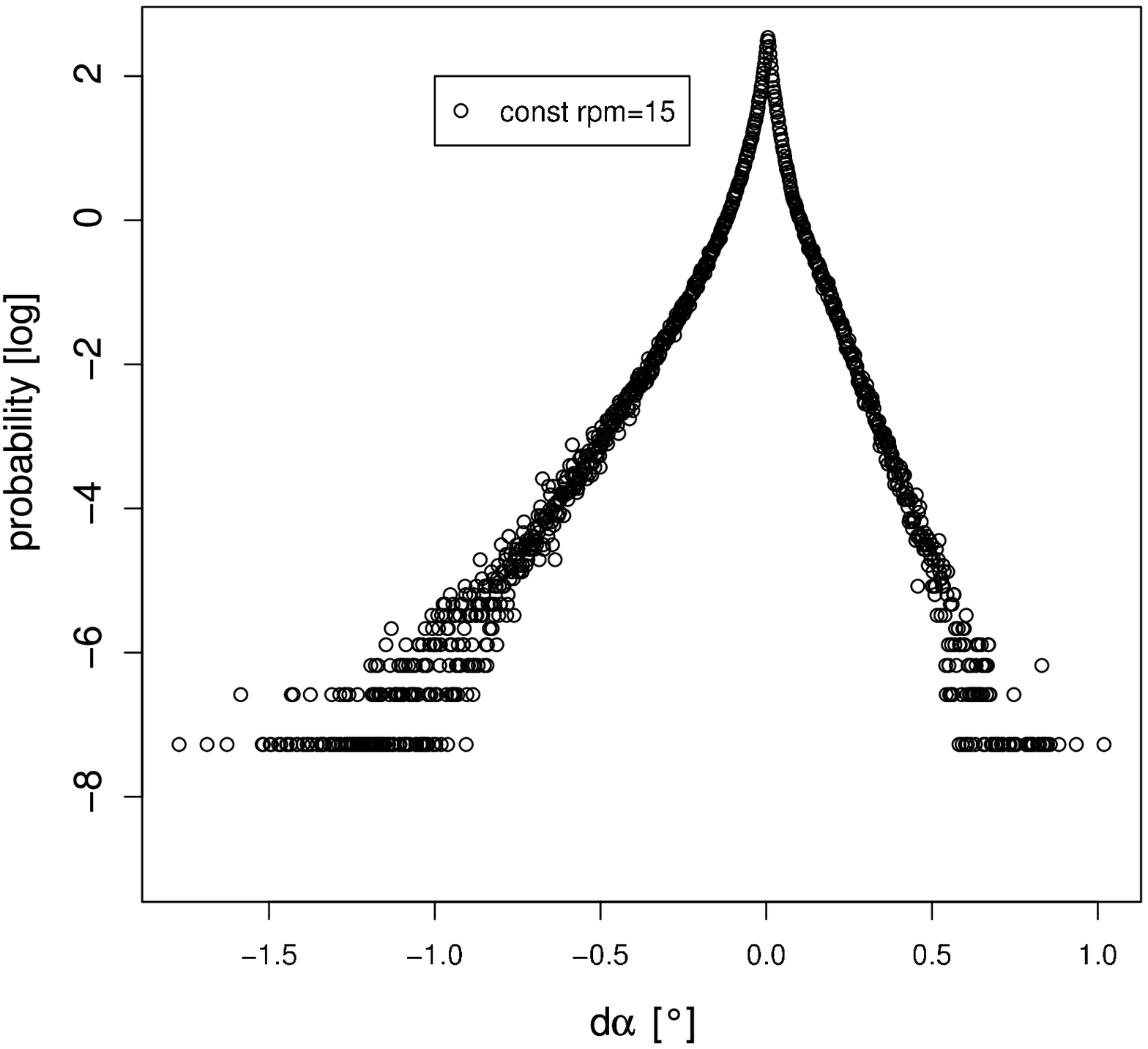}
 \put(-210,150){c)} &
\epsfxsize=2.78in
\epsffile{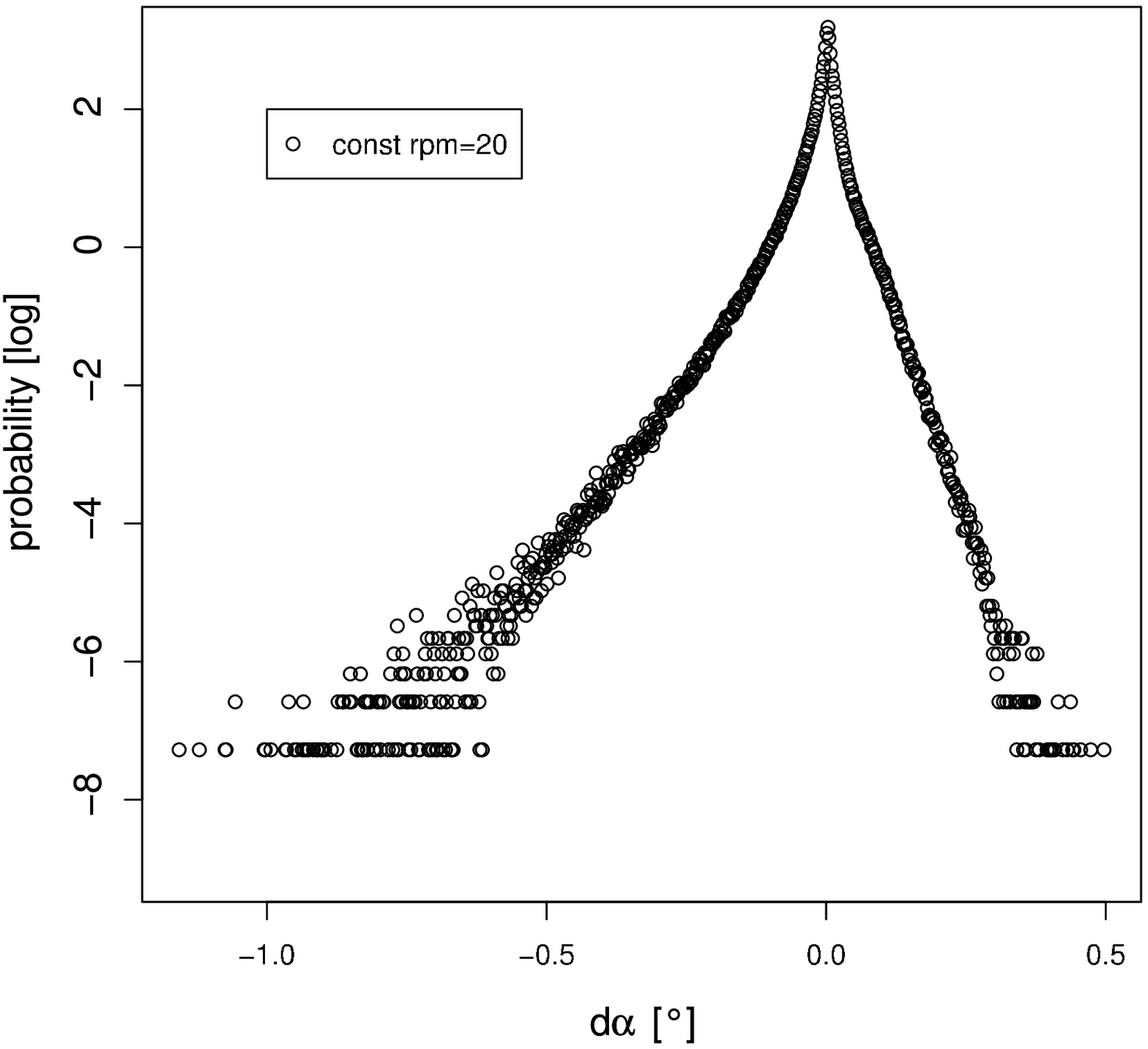}
 \put(-200,150){d)} \\ [-0.3cm]
\end{array}$
\end{center}
\caption{Pdf of the changes in AOA due to change in wind direction within 2 s. a) depicts the case a) of the turbine with a constant tip speed ration. The figures b), c) and d) show the pdfs for the case b) model rotating at a constant speed of $n_{rpm}=10$, 15 and 20 rpm respectively.}
\label{yawangle}
\end{figure}

In the following we analyze the effect of the changes in wind direction on the changes in angle of attack.\\
We regard the maximum changes in AOA under the following conditions: The wind speed is expected to be constant, the averaging period $\tau_a=2$ s and $\lambda_T=7$ for case a) and $n_r=10,15,20$ rpm for the case b) model. Thus for the $n_r=15$ case a rotor blade would perform half a revolution.\\
Using these assumptions the model equations (\ref{tipspeedall}) and (\ref{constspeedall}) change to

\begin{eqnarray}
\Delta\alpha = \arccos \left( \frac {  \frac{\lambda_T r}{R} - \frac{2}{3}\sin(\Delta \phi)}{\sqrt{( \frac{\lambda_T r}{R})^2+(\frac{2}{3})^2-\frac{4}{3}\frac{\lambda_T r}{R}\sin(\Delta \phi)}}\right)-\arctan\left(\frac{2}{3}\frac{1}{\frac{\lambda_T r}{R}}\right),
\label{tipspeedyaw}
\end{eqnarray}
for turbines with constant tip speed ratio and to\\
\begin{eqnarray}
\Delta\alpha = \arccos \left( \frac { u_{rpm}- \frac{2}{3}\overline u(t)\sin(\Delta \phi)}{\sqrt{u_{rpm}^2+(\frac{2}{3}\overline u(t))^2-\frac{4}{3}\overline u(t) u_{rpm}\sin(\Delta \phi)}}\right)-\arccos\left(\frac{u_{rpm}}{\sqrt{u_{rpm}^2+(\frac{2}{3}\overline u(t))^2}}\right)
\label{constspeedyaw}
\end{eqnarray}
for a turbine at a constant rotational speed.\\
Using $\Delta \phi(\tau_a)$ we obtain the pdfs for the changes in AOA presented in fig. \ref{yawangle}.
%

The pdfs for the AOA in fig. \ref{yawangle} show a remarkable difference between the case of a turbine running at constant tip speed ratio and the turbine running at constant rotational speeds. Due to the constant ratio between the wind speed and the rotational speed the pdf of the case a) model shows a distribution limited towards positive deviations. While the pdfs for the case b) model show an intermittent curve, which is strongly asymmetric. The base functions (\ref{tipspeedyaw}) and (\ref{constspeedyaw}) lead to a higher scatter toward negative angles in all cases. As a result all pdfs in the graphs have a negative skewness and very high values for the kurtoses given in table \ref{yawtab}. Comparing the characteristics of the pdfs in table \ref{yawtab} to the values of the pdfs obtained for the complete wind field in table \ref{tabaoages}, we conclude that the influence of the changes in wind direction on the changes in AOA are small.\\
\begin{table}
\centering
\begin{tabular}{|c|c|c|c|}
\hline
{\bf Type} & {\bf Skewness} & {\bf Kurtosis}& {\bf $\sigma$}\\
\hline
$\lambda_T=7$ & -1.7 & 7.5&0.28$^\circ$\\
$n_{rpm}=10$ & -0.9 & 8.2 &0.23$^\circ$\\
$n_{rpm}=15$ & -1.6 & 10.9&0.11$^\circ$\\
$n_{rpm}=20$ & -2.2 & 14.1&0.06$^\circ$\\
\hline
\end{tabular}
\caption{Characteristics of the pdfs of $\Delta\alpha$ due to change in wind direction}
\label{yawtab}
\end{table}
\\

\section{Deeper analysis of the statistics of $\Delta\alpha$}
To develop models for the fluctuations in the AOA, the rate of the changes over time have to be regarded. This helps to derive a model for the turbulence of the changes in the AOA and as a result realistic estimations of extreme values. Also the rates of $\Delta\alpha$ over time is of major interest for dynamic stall modelling. Therefore the fluctuation of an estimated AOA at different time scales were calculated. To gain a quantative image of the magnitude of the time scales of fluctuations in angle of attack, instantaneous changes in the AOA without averaging have to be regarded. Therefore here a modified (case c)) version of equation (\ref{constspeedall}) is being used for the further calculations. Thus  only a comparison between the current wind and the one at a time different $\tau$ is done, so that $\overline \phi(t)$ and $\overline u(t)$ have to be replaced by  $\phi(t)$ and $u(t)$ in equations (\ref{sindt}) and (\ref{constspeedall}) respectively. The time scale was henceforth varied from 10 to $\frac{1}{3}$ Hz.\\

\begin{figure}[htbp]
\begin{center}
$\begin{array}{c@{\hspace{0.3in}}c}  
 \multicolumn{1}{l}{\mbox{\bf }}  &
 \multicolumn{1}{l}{\mbox{\bf }} \\ [-0.5cm]
\epsfxsize=2.7in
\epsffile{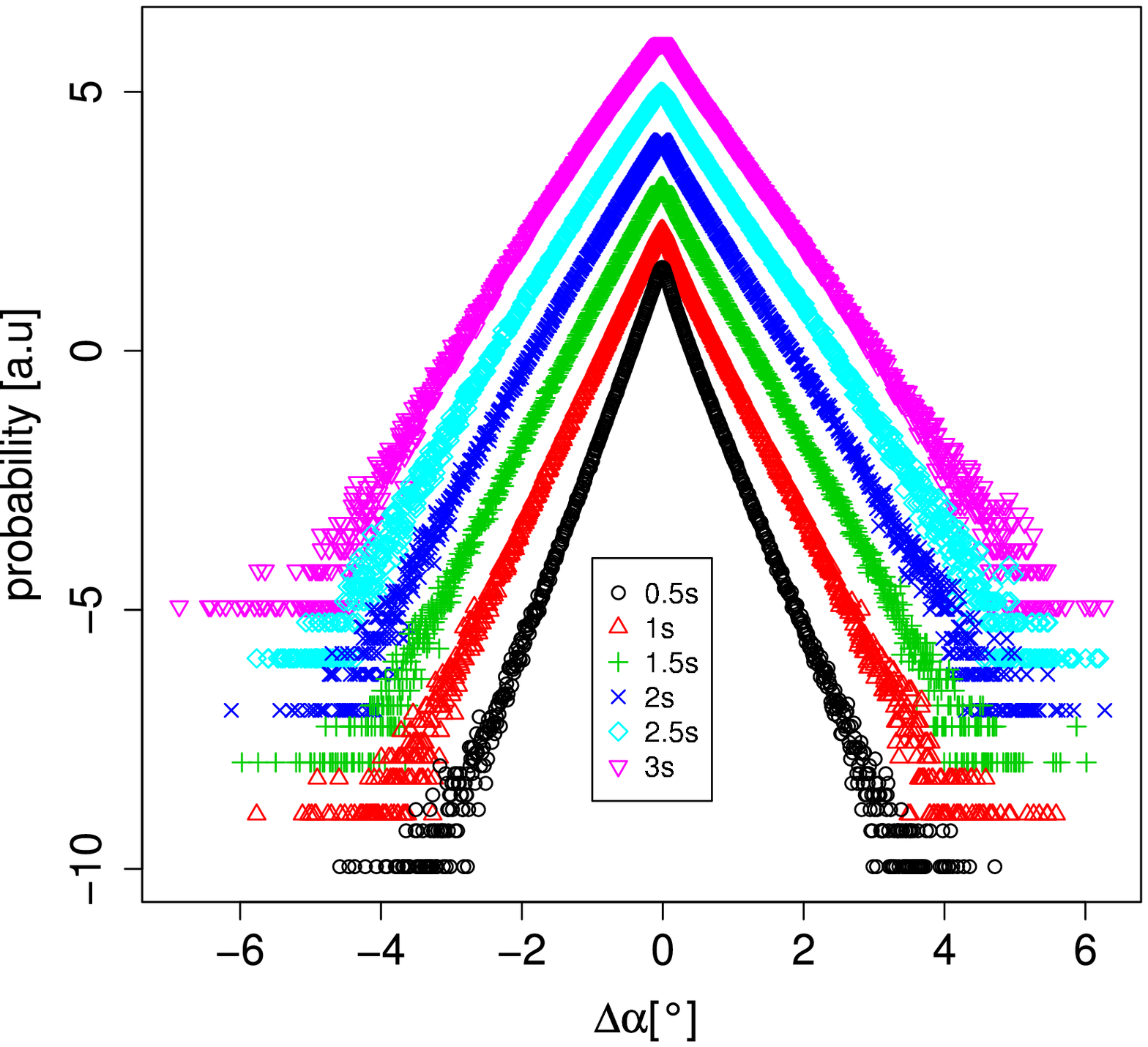}
 \put(-200,150){a)}  & 
\epsfxsize=2.7in
\epsffile{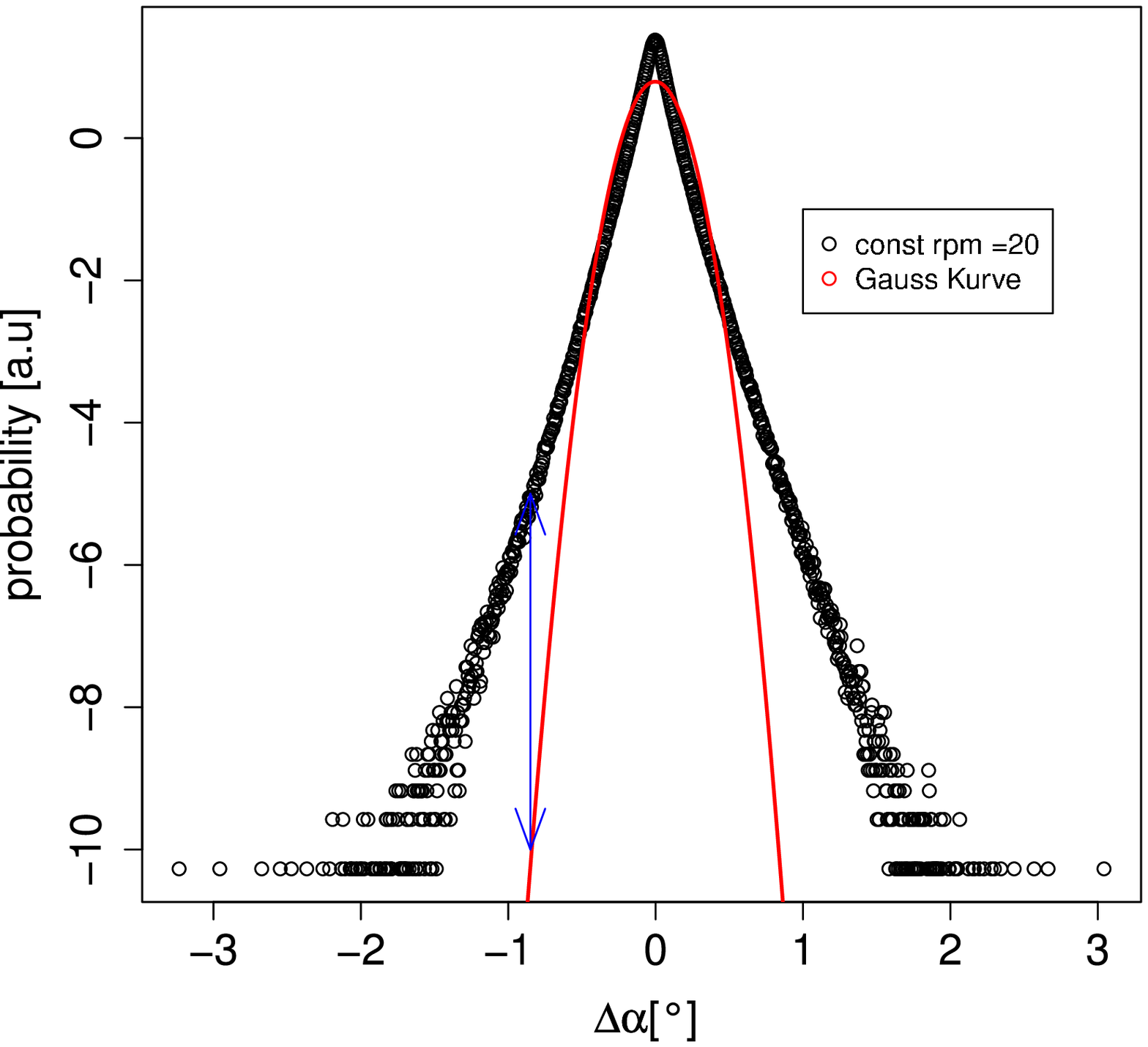}
\put(-200,150){b)}\\ [1.0cm]
\multicolumn{1}{l}{\mbox{\bf }}  &
\multicolumn{1}{l}{\mbox{\bf }}   \\ [-1cm]
\epsfxsize=2.7in
\epsffile{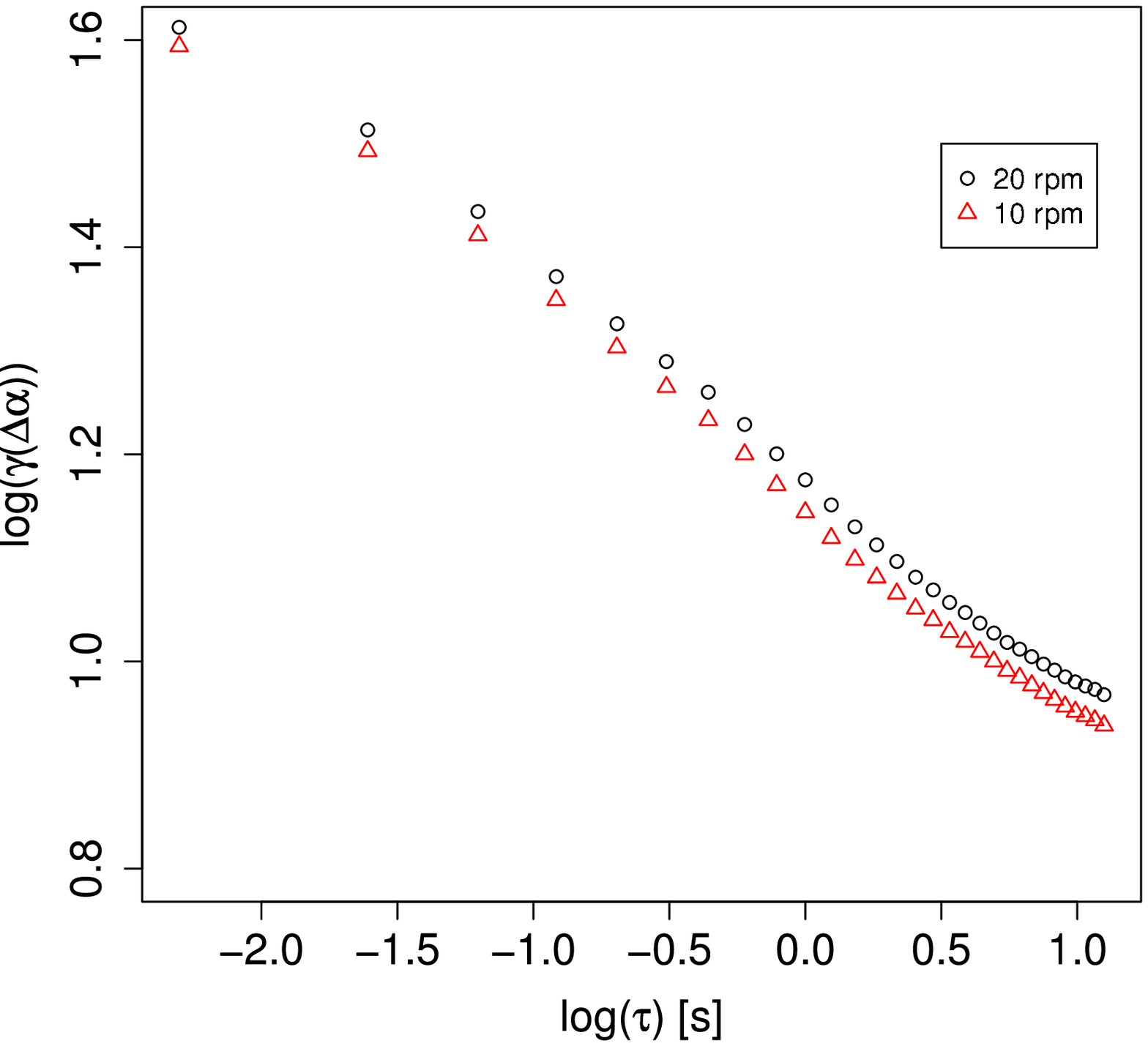}
 \put(-200,150){c)} &
\epsfxsize=2.7in
\epsffile{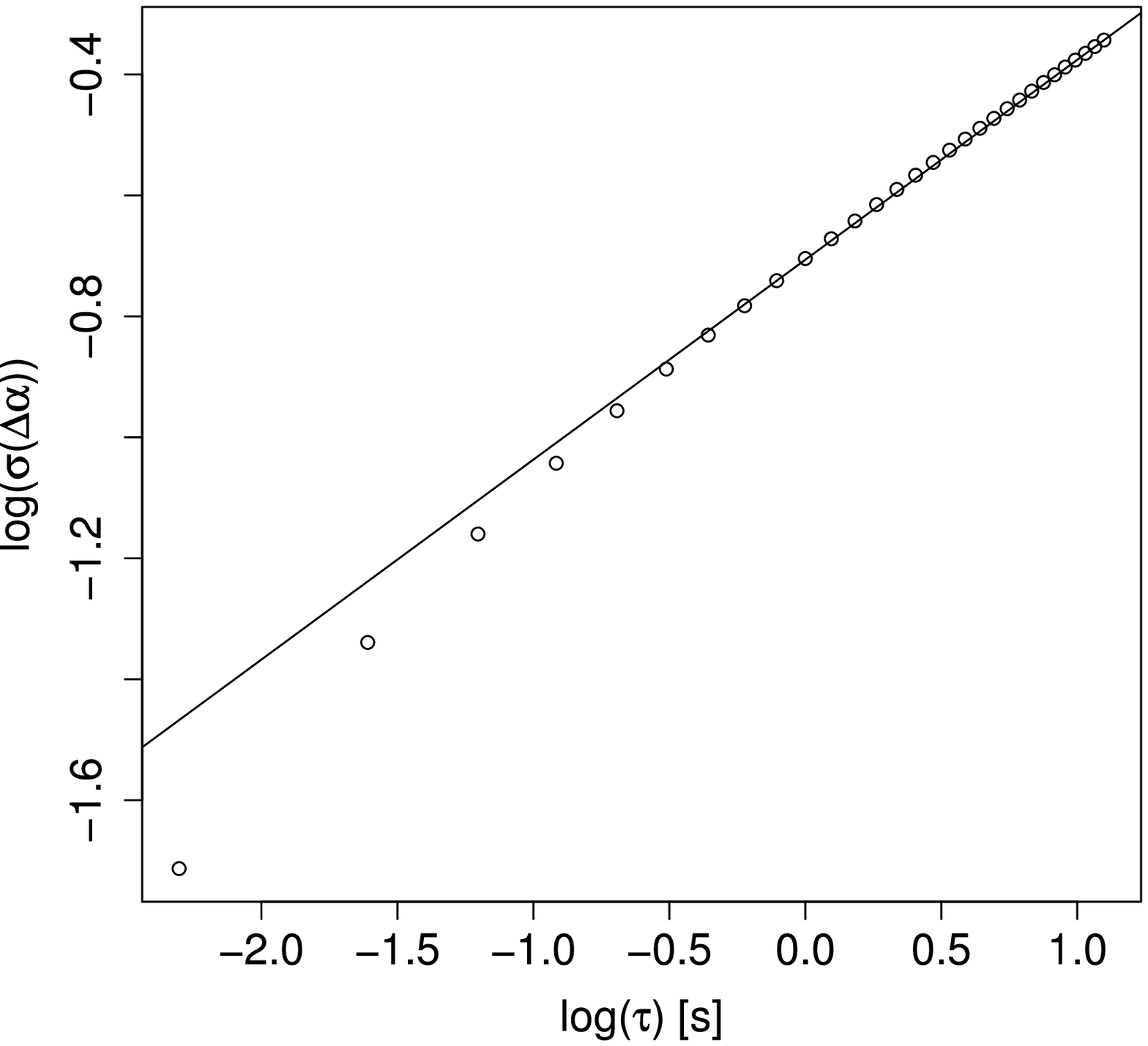}
 \put(-200,150){d)} \\ [-0.3cm]\end{array}$
\end{center}
\caption{The pdfs for different time scales are given in a). As expected $\Delta\alpha$ tends to be broader for longer time scales. The pdfs remain nevertheless intermittent all the time. Fig. b) depicts the pdf of the $\frac {1}{10}$ s time scale against a Gaussian curve. Both figures a) and b) are calculated for a fixed rotation of $n_{rpm}=20$ rpm. The rate of the intermittency over the time scale is given by the kurtosis in fig. c) for fixed rotations of $n_{rpm}=10$ rpm and $n_{rpm}=20$ rpm respectively. Fig. d) shows the broadening of pdfs by its standard deviation and the time scale $\tau$. The line in d) depicts the slope of a linear function of $\frac{1}{3}$ for the case of $n_{rpm}=20$.}
\label{aoadiff}
\end{figure}

\begin{figure}[htbp]
\begin{center}
$\begin{array}{c@{\hspace{0.3in}}c}  
 \multicolumn{1}{l}{\mbox{\bf }}  &
 \multicolumn{1}{l}{\mbox{\bf }} \\ [-0.5cm]
\epsfxsize=2.7in
\epsffile{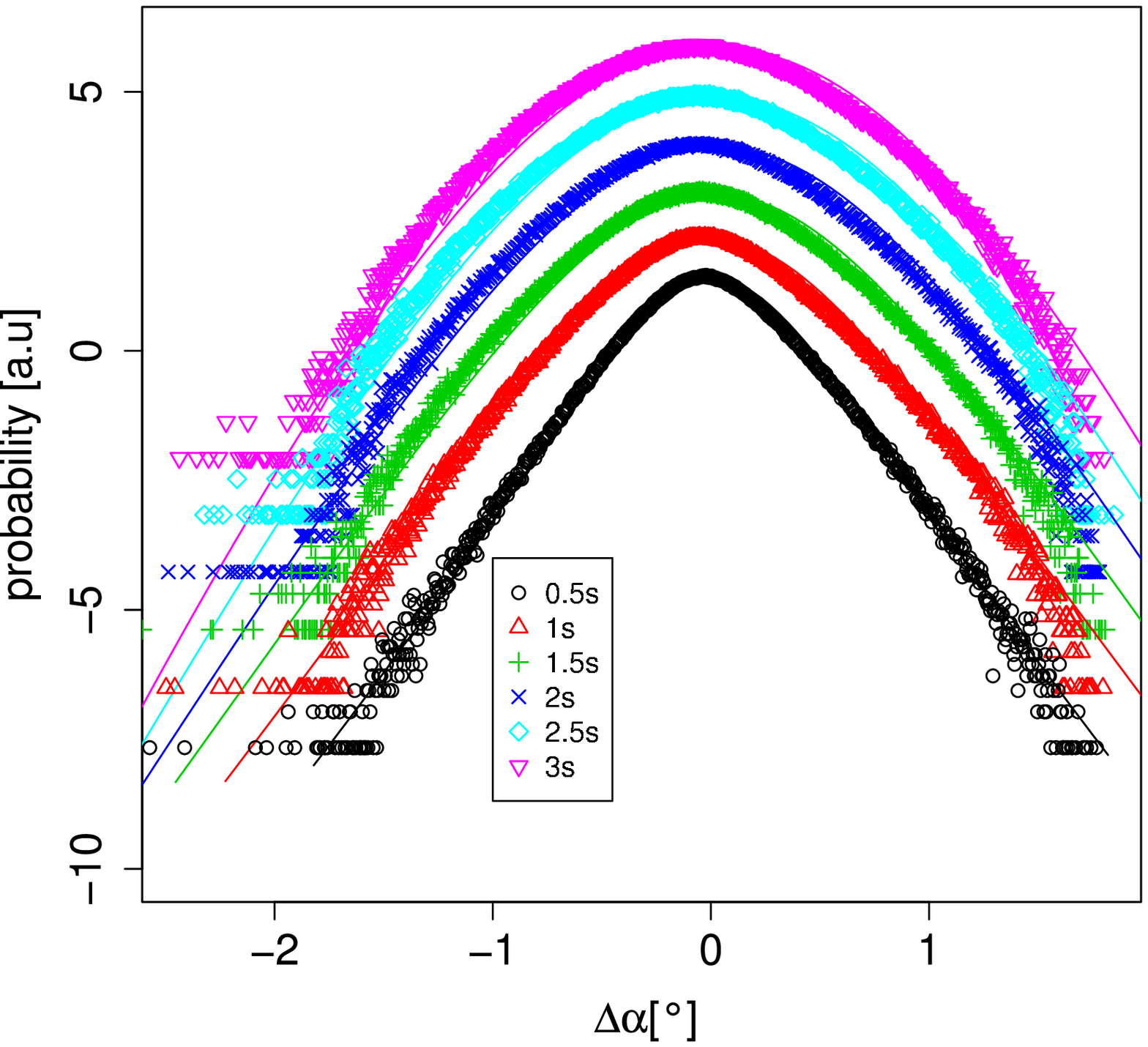}
 \put(-200,150){a)}  & 
\epsfxsize=2.7in
\epsffile{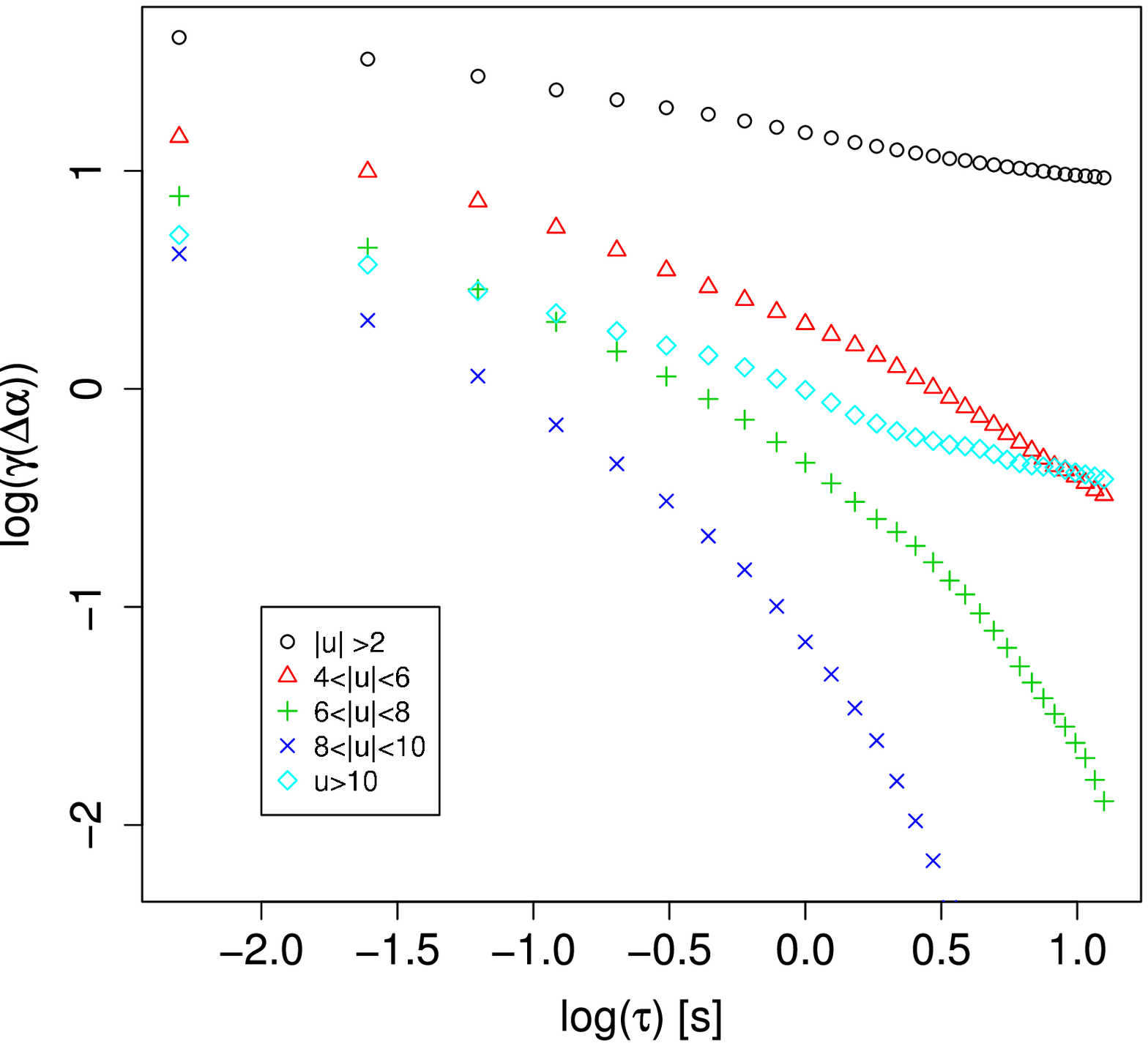}
\put(-200,150){b)}\\ [1.0cm]
\end{array}$
\end{center}
\caption{Looking at pdfs of $\Delta\alpha$ under certain wind speed conditions. In a) pdfs for different time scales under the condition of $6 \leq |u| \leq 8$ m/s show increasingly Gaussian shapes the larger the time scale gets. This is quantified by the plot of the kurtoses in b) for different conditions. Where the original curve shows way larger values.}
\label{cond}
\end{figure}
\begin{figure}[htpb]
\begin{center}
$\begin{array}{c@{\hspace{0.3in}}c}  
 \multicolumn{1}{l}{\mbox{\bf }}  &
 \multicolumn{1}{l}{\mbox{\bf }} \\ [-0.5cm]
\epsfxsize=2.7in
\epsffile{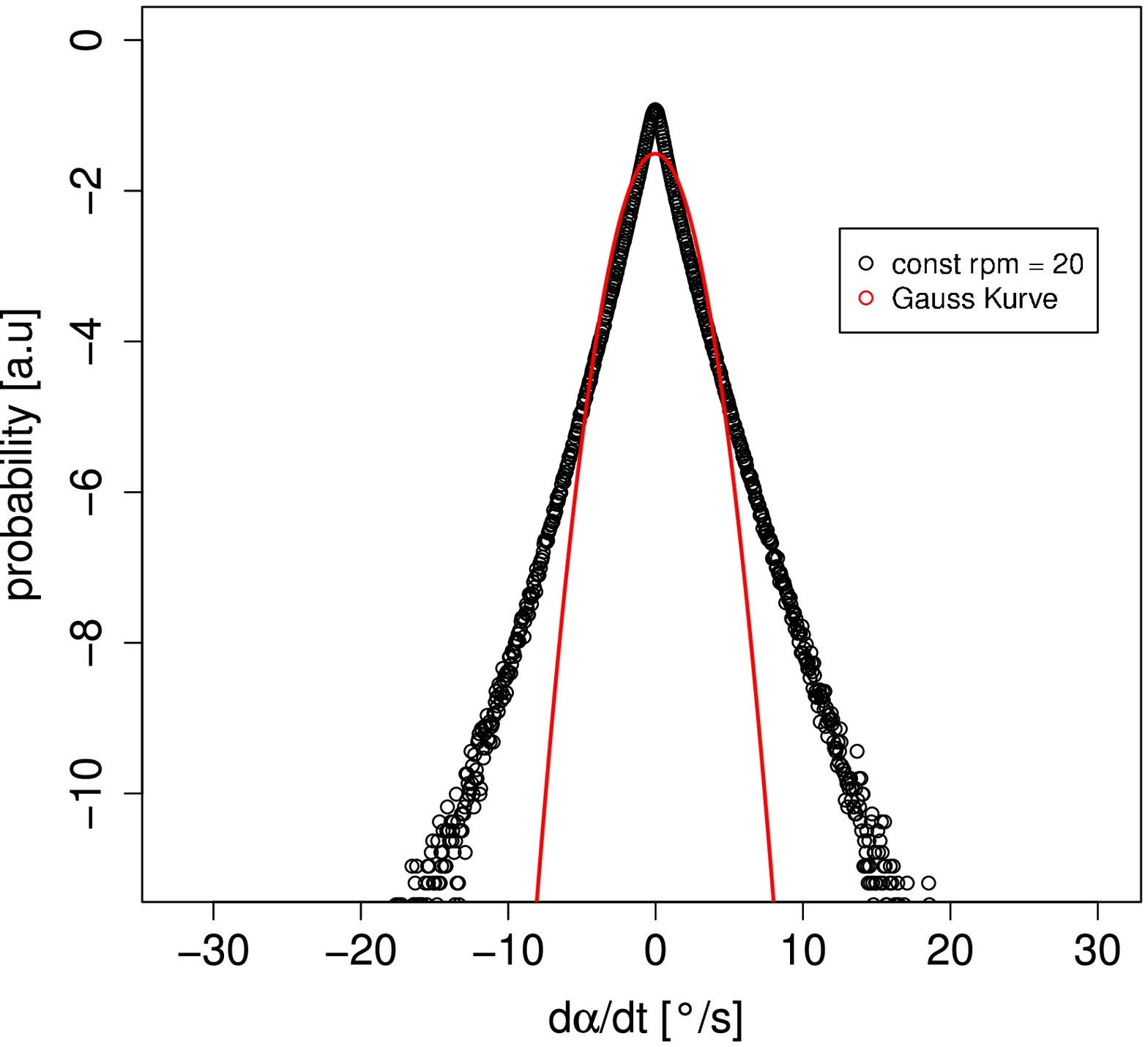}
 \put(-200,150){a)}  & 
\epsfxsize=2.7in
\epsffile{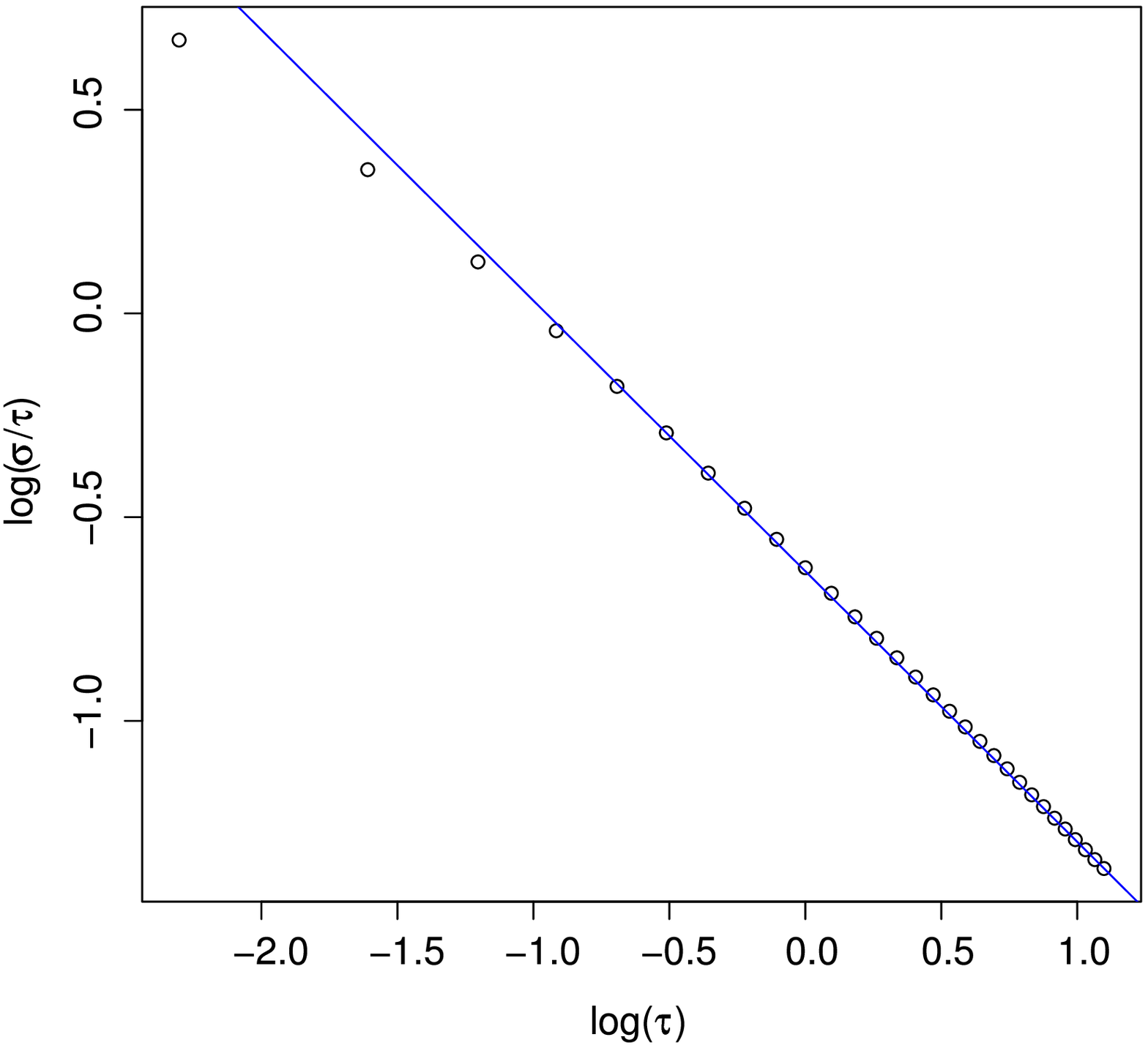}
\put(-200,150){b)}\\ [1.0cm]
\end{array}$
\end{center}
\caption{Probability density function of $\Delta\alpha/\tau$ for $\tau=10$ Hz graphed with a Gaussian distribution for comparison, giving an idea for $d\alpha/dt$ a).  The maximum rates and the rates for the standard deviation of the distribution for different $\tau$ is given in b).}
\label{daodt}
\end{figure}

It can be seen in fig. \ref{aoadiff} a) a main portion of the changes in $\Delta\alpha$ appear to evolve at short time periods. For the characterization of the distributions we plotted the kurtosis over the time scales in fig. \ref{aoadiff} c). It can be observed that the kurtosis of the pdfs shrinks only slowly for time scales of 10 - $\frac{1}{3}$ Hz. As an example of the deviation from a Gaussian distribution fig. \ref{aoadiff} b) depicts the distribution for a time scale of $\tau=\frac{1}{10}$ s.\\
The knowledge of the pdfs make reconstructions of such distributions by analytical models in case of wind speed distributions possible \cite{boettcher03}. If such models are applicable also to the change in AOA $\Delta\alpha$, they can further be used to optimize models for dynamic stall on wind turbines. Castaing et. al. proposed to fit the pdfs of $\Delta u$ to a function that would for $\Delta\alpha$ look like
\begin{eqnarray}
P(\Delta\alpha_{\tau})=\frac{1}{2\pi \lambda}\int_0^{\infty}exp\left(-\frac{\Delta\alpha_{\tau}^2}{2\sigma^2}\right)exp\left(-\frac{ln^2(\sigma/\sigma_0)}{2\lambda^2}\right)\frac{d\sigma}{\sigma^2}
\label{Castaing}
\end{eqnarray}
where $\sigma_0$ is the most probable variance of $\Delta\alpha$ \cite{castaing90}. By determining the $\lambda$ we thus gain the analytical fit function $P(\Delta\alpha_{\tau})$ of the pdf for the time scale $\tau$. Since $\lambda^2$ can be taken from
\begin{eqnarray}
\lambda^2\propto ln(\gamma),
\label{lambda-2}
\end{eqnarray}
the knowledge of the different statistical properties as $\sigma(\tau)$ and $\gamma(\tau)$ is essential, they describe the pdfs at different conditions. In the following they shall be described in more detail to give an overview over the characteristics of the pdfs.\\
The distribution of the pdfs given in fig. \ref{aoadiff} a) only broaden at a mediocre rate as the time scale increases. In fig. \ref{aoadiff} d) the double logarithmic plot the standard deviations $\sigma$ of the pdfs of $\Delta\alpha$ over the different time scales are shown. For time scales $\tau \geq 0.6$ s the broadening of the standard deviation can be well described by a power law of $\sigma \approx \tau^{\frac{1}{3}}$. If we assume the Taylor hypothesis to be valid in that range 
\begin{eqnarray}
\Delta\vec{r} = <\vec{u}>\Delta t,
\label{taylor}
\end{eqnarray}
the slope of the curve in fig. \ref{aoadiff} d) coincides with the turbulence theory of Kolomogorov from 1941 for isotropic and homogeneous turbulence \cite{Kolm41} in case of the second structure function, stating
\begin{eqnarray}
S^{n}_{u}(r)\propto r^{\frac {n}{3}}.
\label{Kol41}
\end{eqnarray}
Here S is the structure function of the n-th order. As the variance is the structure function of the 2-nd order, $\sigma \propto r^{\frac {1}{3}}$ could be expected for Gaussian turbulence. So even though the overall pdfs for the change in angle of attack for time scales $0.6 \leq \tau \leq 3.0$ s show a non Gaussian behavior, the relations for the structure function for these time periods behave according to \cite{Kolm41} like a Gaussian distribution.\\
A possible reason for this could be, that the overall distribution consists of a superposition of many mainly almost Gaussian distributions. In fig. \ref{cond} the pdfs and the kurtoses over time scales are shown for wind speeds under the restricting condition $6 \leq |u| \leq 8$ m/s. Already the pdfs can clearly be identified to be a lot closer to a Gaussian shape then it was the case for the non-conditioned wind speed in fig. \ref{aoadiff}a). Though $\tau \le 2$ s the kurtosis of the pdfs is still positive, it is a lot smaller than for the over all wind field. This explains why obviously equation (\ref{Kol41}) is in our case still valid. \\
As could be seen in fig. \ref{aoadiff}, for time scales $\tau <0.6$s the relation of equation (\ref{Kol41}) does not hold anymore. This is also reflected for the conditioned pdfs in fig.\ref{cond} for small time scales $\tau \leq 0.5$ s. The kurtoses do not decrease or approach a value close to $\gamma=0$.\\
Using this background it is a reasonable attempt to fit a reconstructed pdf to the data to gain an analytical function  with equation (\ref{Castaing}) for the distribution. Such a fit has been done for the conditioned pdfs in fig. \ref{cond} a). The pdfs give to a good extend the characteristics of the changes in the angle of attack at a certain position of a wind turbine blade within a given period of time.\\
Since for the calculation of dynamic stall not only the magnitude, but also the time in which changes in angle appear are important, the rates of the changes in time are being presented here. It is straight forward to calculate the distribution of $d\alpha/dt$ for a sampling rate of $\tau=10$ Hz. Fig. \ref{daodt} a) shows the distribution for $\Delta\alpha/\tau$ for a rotation of $n_{rpm}=20$ rpm. In fig. \ref{daodt} the rates are plotted for the standard deviation of the pdfs and the maximum values for different $\tau$.\\

\vfill

%
%
\section{Discussion and Conclusion}
Inspired by the strong simplifications in the modeling of dynamic stall, we have shown an analysis of the statistical properties of the changes in the angle of attack estimated from the wind field. After a short analysis to characterize the wind field, we proposed a simple comprehensive model to estimate the changes in the angle of attack on a turbine blade at a certain rate of time. Basic turbulence characteristics from the wind field were thus transformed into characteristics which could be used for the modeling of dynamic stall on wind turbine blades.\\
The statistical properties of the wind speed and the wind direction showed the typical intermittent behavior. Even though the latter had a tendency to be less intermittent. Both changes in wind cause changes in the angle of attack on blades of a rotating turbine. Such changes therefore reflect the distributions of the changes in wind and showed in general also an intermittent distribution. Different cases were regarded in the study.\\
Two model cases were used to calculate the changes in AOA: case a) for constant tip speed ratio and case b)  constant rotational speed. For both cases the changes in AOA with a moving average at a time scale $\tau_a=2$ s were analyzed. In case a) the adaption of the rotational speed to a moving average of the wind speed caused the pdfs of the changes in AOA to be less intermittent than the ones for turbines running at a constant rotation rate.\\
To analyze the asymmetry in the pdfs, the same cases have been evaluated for the instance of a change in wind direction only. Here pdfs for the changes in AOA show a strongly skewed behavior. However for the over all pdfs of the changes in AOA caused by wind speed and wind direction changes, it could be seen that the changes in wind speed dominate the effects on the changes on the AOA.\\
Since the aim was to find models for the changes in AOA that could be used for dynamic stall modeling the turbulent structure of the changes have been further analyzed. It was found for turbines running at a constant speed, that for time scales $\tau \geq 0.6$ s the standard deviation of the pdfs broadens by a power law $\approx \tau^{\frac{1}{3}}$. From this was concluded that the intermittent distributions over the complete time consist most likely of a superposition of mostly Gaussian distributions. For time scales $\tau < 0.6$ s this is not true anymore. So for AOA changes at time scales of $\tau \leq 0.5$ s larger values of the kurtoses and smaller values of the standard deviation are to be expected - indicating intermittent, non-Gaussian fields. Due to the small time scales, this is expected to lead to local aerodynamic effects on the  blade as dynamic stall and will not happen on scales to which an existing controlling system would be able to react.\\
Since the main interest was to find characteristic distributions for the dynamic stall modeling the resulting conditioned pdfs for the changes in AOA have been fitted. The results showed slight deviations caused by the changes in wind direction. However the major characteristics were grasp very well. This makes new approaches for the modelling of dynamic stall for the optimization of wind turbine design possible.\\

\subsection*{Acknowledgements}
We would like to thank my (former) Co-Workers Julia Gottschall, David Kleinhans, Robert Stresing und Tanja M\"ucke for their support and advice on the paper.
%
%
%
 
\bibliographystyle{aiaa}
\bibliography{BSwejp}

\end{document}